# OUTSTANDING FRAMEWORK FOR SIMULATING AND GENERATING ANCHOR TRAJECTORY IN WIRELESS SENSOR NETWORKS

Abdelhady Naguib [1, 2]

[1]Department of Computer Science, College of Computer and Information Sciences, Jouf University, Sakaka, Saudi Arabia
[2]Department of Systems and Computers Engineering, Faculty of Engineering, Al-Azhar University, Cairo, Egypt

## ABSTRACT

*This paper proposes a framework that has the ability to animate and generate different scenarios for the mobility of a movable anchor which can follow various paths in wireless sensor networks (WSNs). When the researchers use NS-2 to simulate a single anchor-assisted localization model, they face the problem of creating the movement file of the movable anchor. The proposed framework solved this problem by allowing them to create the movement scenario regarding different trajectories. The proposed framework lets the researcher set the needed parameters for simulating various static path models, which can be displayed through the graphical user interface. The researcher can also view the mobility of the movable anchor with control of its speed and communication range. The proposed framework has been validated by comparing its results to NS-2 outputs plus comparing it against existing tools. Finally, this framework has been published on the Code Project website and downloaded by many users.*

## 1. INTRODUCTION

Wireless sensor networks have benefits for different areas of applications, such as military, health care, resident applications, tracking, disaster prediction and recovery, and IoT (the Internet of Things). On the other hand, there are many challenges they face, such as restricted computing and processing, energy consumption, and storage capability [1]. WSN localization is the process of finding the location of a sensor node in the deployment field. Localization plays an important role in various WSN applications. It is useless to collect data without knowing the location of the source or event [2]. Many localization mechanisms have been proposed in recent years, but one of the most promising techniques is called static path planning algorithms, which use a single movable anchor node equipped with a GPS unit. The movable anchor node follows a predefined static path, broadcasting its current position periodically to unknown position sensor nodes. Unknown position sensor nodes can use localization algorithms such as range-free or range-based [3] to determine their location using information gathered from the movable anchor node.

Communication networks have become more enhanced and complex, which requires designing a modern and improved model to help in making an accurate analysis and in-depth understanding of the performance of these systems [4]. The simulation process is necessary to develop any modern model or algorithm before it is implemented in the real world. Therefore, a simulation process is carried out to verify the suitability of the system to be implemented for the environment under implementation and its operability according to the functions desired from it. Before the virtual implementation of the system, simulation is performed in order to check for severe weaknesses or other design problems that may occur in the future [5]. On the basis of the type of application area in wireless sensor networks, simulation can be classified into continuous simulation, discrete event simulation (DES), deterministic simulation, parallel simulation, stochastic simulation, and hybrid simulation [6]. Below is a detailed description of these categories:

- **Continuous simulation:** Continuous simulations can be described by a set of constantly changing dependent variables. Continuous simulations deal with the modeling of physical events (processes, behaviors, and conditions) [7] used in differential equations that describe the physical processes.
- **Discrete event simulation:** This type of simulation is applied to systems in which events occur over discrete time periods because no change is expected between events, so every change observed in the system is equivalent to the occurrence of a specific event [8].
- **Stochastic simulation:** There are some systems in which variables change randomly as probabilities change and which have several possibilities for evolution. Applications used for this type of simulation are to simulate traffic flow in communications networks or to study climate changes [9].
- **Deterministic simulation:** A future event can be calculated accurately, without interfering with randomness, using deterministic simulation which has a previously known input and a unique set of outputs [10].
- **Hybrid simulation:** This type of simulation is a hybrid of the two types of simulations defined above, so it is considered more efficient for testing the target system because it uses more than one simulation model to evaluate the model [11].

The rest of this paper is organized as follows: Section 2 presents related works. Section 3 describes the process of framework analysis & design. Section 4 illustrates the framework interface design process. Section 5 presents the results verification and discussions. Finally, Section 6 provides the conclusion of this paper.

## 2. RELATED WORKS

This section presents static path panning algorithms that the proposed framework supports. The work presented in [12] proposes three different types of static path planning algorithms. The first algorithm is SCAN, where the movable anchor node starts moving vertically along with the monitoring area, and then moves horizontally by a distance equal to the trajectory resolution. The second algorithm is DOUBLE-SCAN, where the movable anchor node scans the deployment field in both directions so as to overcome the collinearity problem of the SCAN algorithm. The third algorithm is HILBERT, where the monitoring area is divided into four square regions connected by four line segments, ensuring that sensor nodes have no chance to receive collinear beacons from movable anchor node. Authors in [13] propose two static path- planning algorithms: CIRCLES and S-CURVES. According to CIRCLES, the movable anchor node follows an arrangement of concentric circles covering the monitoring area except the four corners. Hence, the second algorithm, called S-CURVES, tries to cover the four corners of the deployment field and avoid collinear beacon transmission. It allows the movable anchor node to move according to the S-shape both vertically and horizontally, like the SCAN algorithm.

Zhen Hu et al. [14] proposed a mobile anchor centroid localization method (MACL) where the movable anchor node moves in a spiral path, broadcasting its current position periodically.

Authors in [15] proposed a localization algorithm with a movable anchor node based on trilateration (LMAT). This method uses a movable anchor that moves according to the trilateration path in the sensing field while broadcasting its current position periodically. Y. C. Lin et al. [16] proposed a localization algorithm based on the triangle grid scan (TGS) method. This method aims to use Wi-Fi technology for anchor movement in an easy and simple manner.

The work in [17] proposes a new, superior method for path planning known as a Z-curve. To get around the issue of collinear beacons, the movable anchor moves along a static path that resembles a Z shape with several z-curves in the deployment region. For two-dimensional WSNs, the authors in [18] suggested a mobile anchor-assisted localization method (MAALRH) based on regular hexagons. A succession of regular hexagons that cross each other in the middle make up the path taken by the movable anchor. The entire sensing field is covered in this manner. A path-planning model called H-Curves was created by Alomari et al. [19] for mobility-assisted localization in wireless sensor networks. A movable anchor's trajectory is made up of numerous H-shaped pathways to get around the issue of collinear beacons.

K. Kannadasan et al. [20] proposed a novel path planning method for movable anchor-assisted localization called M-Curves. The authors optimized the localization system by using the Dolphin Swarm Algorithm (DSA) [21] so as to increase localization accuracy. Abdelhady Naguib and S. Ali [22] proposed a new static path-planning approach used in wireless sensor networks called SQUARE_SPIRAL, which enables a movable anchor to cover the whole deployed area to get high coverage. The trajectory of the movable anchor node follows a square spiral shape to decrease the number of collinear beacons and increase localization accuracy. Authors in [23] proposed a novel path planning method for movable anchor-based localization called Nested Hexagon Curves (NHexCurves). The trajectory of the movable anchor node based on drawing a pattern consisting of N-HexCurves so as to overcome the problem of collinear beacons using weighted centroid localization (WCL) [24] and accuracy priority trilateration (APT) [25] localization.

In this article, an outstanding framework for simulating anchor movement based on different static path-planning models is proposed. The proposed framework is an open-source tool that can be downloaded from the Code Project website, allowing researchers to change it to fit different scenario models or modify to it to make it work with other path planning mechanisms. The proposed framework has been tested with the well-known simulator NS-2, and the movable anchor's movement has been confirmed by running the same motion scenario files on the Network Animator (Nam) tool. The proposed framework has the following features: generating external text files storing the coordinates of movable anchor trajectories according to different static path planning mechanisms, animation for the movement of the movable anchor, and running of already-generated scenarios for the sake of comparison between different trajectory models.

## 3. FRAMEWORK ANALYSIS & DESIGN

This section is concerned with developing an object-oriented model of the proposed framework. The object class hierarchy of the proposed framework is shown in Figure 1 and consists of the following main classes: The first class is **User_Interface**, which has many attributes like *total sensors*, which represents the total number of sensors plus one anchor node. The other attribute is *simulation Time*, which represents the duration of the simulation. This class has two main operations: The first one is **network_Paint**(), which is responsible for painting objects on the specified GUI painting area. The other operation is **timerUpdate_Tick**(), which represents the elapsed real time.

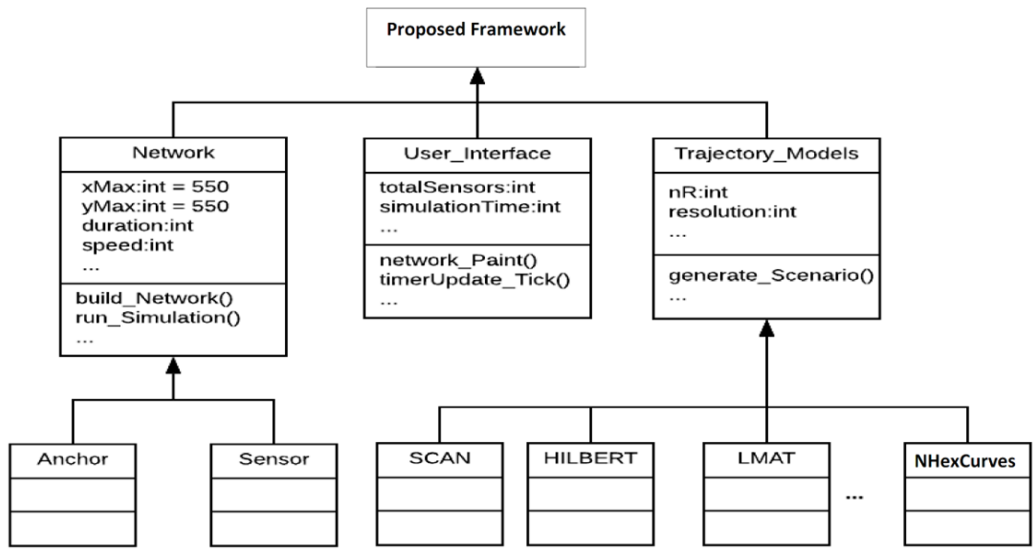

Figure 1. Framework Class Hierarchy

The second class is **Network**; this class contains all the attributes and operations needed for the wireless sensor network. The first main attribute is *xMax*, which represents the width of the deployment field in the x direction and has a default value of 550 *m*. The second one is *yMax*, which refers to the height of the deployment field in the y direction; it is also the default value, settled at 550 *m*. The other attribute is *duration*, which represents simulation time in seconds. The other attribute that controls the speed of the movable anchor is *speed* (*m/sec*). There are many operations associated with this class; The first main operation is **build_Network**(), which is responsible for setting the control data from the user interface. The second main operation is **run_Simulation**(), which controls the core processes for running the simulation.

The **Network** class has two sub-classes: **Anchor** and **Sensor**. As shown in Figure 2, the **Anchor** class defines a number of attributes that hold information about anchors, including their *ID*, (*x*, *y*) position, *Radius*, which represents the communication range of the anchor, and *InitialEnergy*, which is a *double* variable that represents the initial energy of the anchor node measured in *joules*. Operations associated with this class are **move(x, y),** which is required for moving the movable anchor. The other operation is **send (m),** which is called periodically when the anchor broadcasts its current coordinates to neighboring sensor nodes.

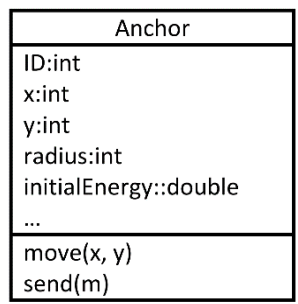

Figure 2. Anchor Class

The third main class of the proposed framework is **Trajectory_Models**; this class represents the static path planning mechanisms used in WSNs. There are many attributes for this class; the first main attribute is *nR,* which refers to the number of horizontal *R* segments needed for determining the size of the monitoring area. The second one is *resolution,* which refers to the length of the *R* segment; thus, According to operations associated with this class like **generate_Scenario()**, it is responsible for writing the coordinates of the movable anchor to an external text file. There are many trajectory models that inherit from the parent class **Trajectory_Models**. Figure 3 represents an example of a child class called **HILBERT**. The first two attributes of this class are *xInitial* and *yInitial,* which represent the initial position of the anchor in *x* and *y* coordinates, respectively. The third attribute is *curveLevel,* which refers to the level (1, 2... n) of the Hilbert trajectory model.

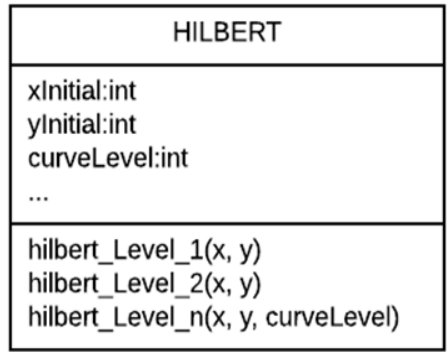

Figure 3. Hilbert Class

There are three operations needed to assign coordinate points to the movable anchor's trajectory. The first one is **hilbert_Level_1(x, y),** which takes two argument variables (*x*, *y*) that represent the initial position of the movable anchor node. The second one is **hilbert_Level_2(x, y)**, which is composed of many level-1 curves. For higher-level Hilbert curves, it is composed of many Level-2 curves that can be modeled by calling **hilbert_Level_n(x, y, curveLevel)**.

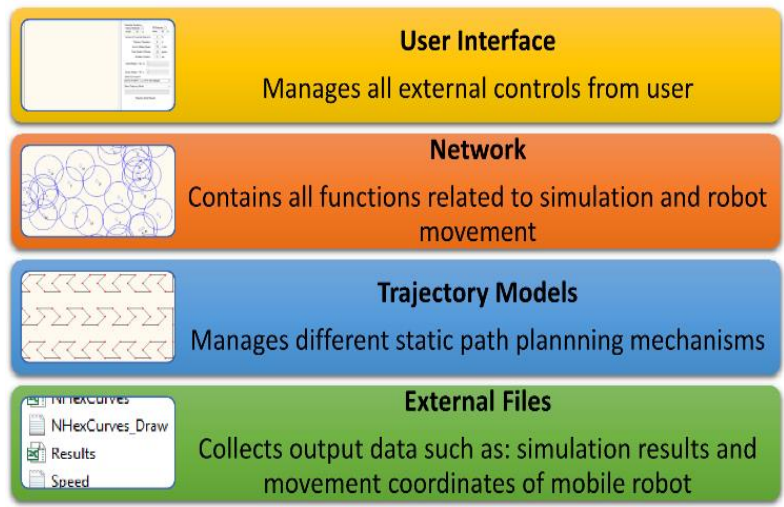

Figure 4. Framework Architecture

The first step of the design process is based on defining the context of the proposed framework. The proposed framework supports multithreading, where the simulation process of the path planning methods runs on a separate thread from the thread controlling the process of the user interface. For better design, each class is encapsulated into a separate reference file. Hence allowing researchers to refer to their own trajectories within the proposed framework. The second step is to design the framework architecture as shown in Figure 4.

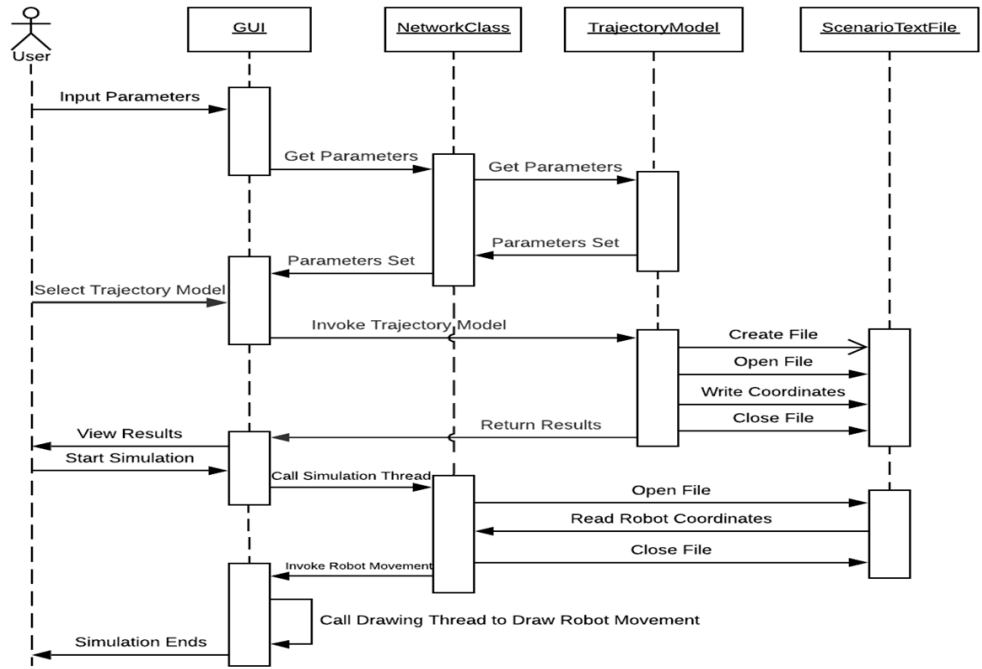

Figure 5. Framework Sequence Diagram

The proposed framework is a simple and flexible system with a layered model consisting of four main parts. The third step in architectural design is to model a system sequence diagram, as shown in Figure 5. The sequence of interactions when the user runs the simulation is as follows:

1) The user should enter the input data that is required for running the simulation.
2) After setting the required parameters, the user selects the required trajectory model, which invokes its object to calculate the path coordinates of the movable anchor node. Then, it will be written to an external scenario file.
3) The results and statistics will appear on the screen and be recorded in an external Excel file.
4) After the user presses the *start simulation* button, the coordinates of the anchor's trajectory will be read by opening the scenario file through the simulation thread.
5) The thread that draws the objects on the animation area runs over and over again while the anchor moves around the sensing field.
6) The simulation stops after the movable anchor node arrives at the end of the trajectory.

## 4. FRAMEWORK INTERFACE DESIGN

The design of the user interface is considered an essential part of the overall software design process. The following are the design decisions that were considered:

1) **User familiarity:** the user interface should consider some concepts in light of the experience of the users who will make the most use of the proposed framework.
2) **Consistency:** The user interface has a consistent design where comparable operations can be invoked in the same way.
3) **Recoverability:** the design of the user interface supports recoverability, so it can recover from errors.

4) **User feedback:** the interface has helpful mechanisms that guide the user when errors occur.

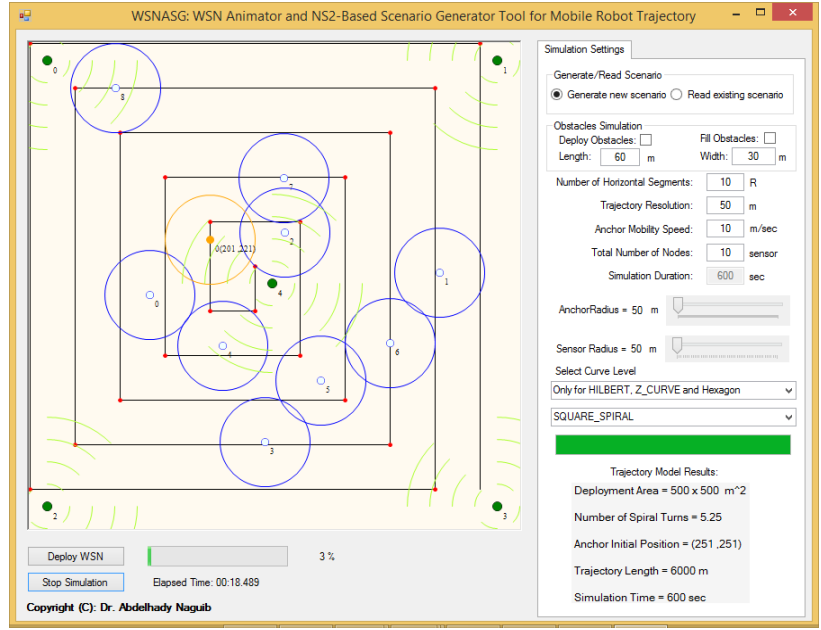

Figure 6. Proposed Framework Interface

Figure 6 presents the main screen of the proposed framework. As shown, the interface is composed of two parts: the first part on the left is the drawing (animation) area where sensor devices and the movable anchor node are deployed. The second part on the right is concerned with simulation settings and a data entry and selection panel where the user can change and select simulation data. The bottom part of the simulation settings panel shows the trajectory model results. The user can select either the *Generate new scenario* or *Read existing scenario* radio buttons. By default, the *Generate new scenario* radio button is selected initially. The user can change the following parameters*: Number of horizontal segments*, *Trajectory resolution*, *Anchor mobility speed, Total number of sensor nodes,* and other settings. After that, the results of the selected trajectory model will be displayed in the lower right corner of the *simulation settings* panel. The displayed results differ according to the chosen trajectory model, but the common results among all trajectory models are as follows: *Deployment area size*, *Total trajectory length,* and *Simulation time*. Afterward, the user can press the *Deploy WSN* button, and the selected trajectory model will be displayed on the monitoring area. Finally, the user can press the *Start Simulation* button to begin a movement of the movable anchor across the drawn trajectory. If the user needs to run an existing scenario, he can click on the *Read existing scenario* radio button.

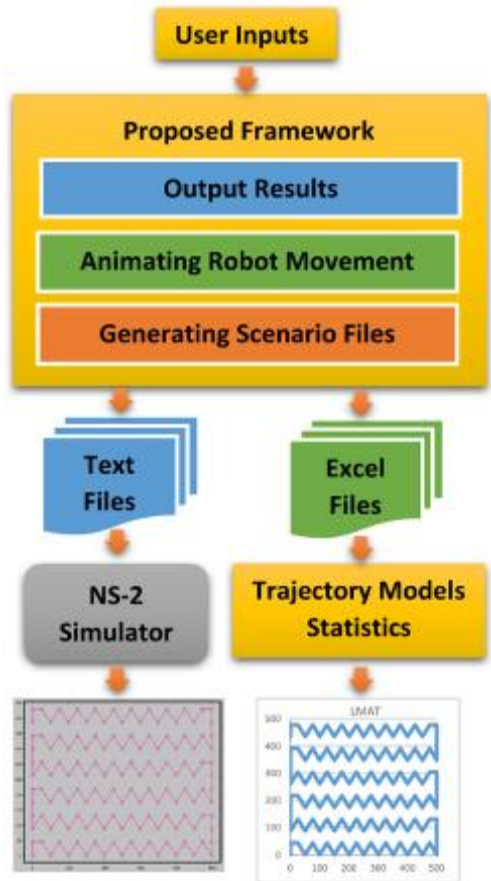

Figure 7. Block Diagram of the Proposed Framework

The block diagram of the proposed framework is shown in Figure 7. The proposed framework can generate simulation results in two ways: the first is by writing the results on the screen of the proposed framework, and the second is via two Excel files. The first Excel file can be used for calculating more statistics about the results gained, and the second Excel file can be used for drawing a directed graph of the selected trajectory model.

```
$ns_ at 0 "$node_(1) setdest 234 301 10.0"
$ns_ at 5 "$node_(1) setdest 201 301 10.0"
$ns_ at 8.3 "$node_(1) setdest 201 201 10.0"
$ns_ at 18.3 "$node_(1) setdest 301 201 10.0"
$ns_ at 28.3 "$node_(1) setdest 301 351 10.0"
$ns_ at 43.3 "$node_(1) setdest 170 351 10.0"
……
```

Figure 8. Generated Scenario File (Sample of Commands)

The proposed framework is capable of generating two scenario files (text files): the first one for storing the topology of the WSN for the purpose of comparing multiple scenarios, and the second

International Journal of Computer Networks & Communications (IJCNC) vol 16, No 6, November 2024

text file for storing the coordinates of the movable anchor's trajectory that can be used as input to the NS-2 simulator. This file contains the commands (shown in Figure 8) needed by NS-2 to simply move the movable anchor node according to the given trajectory model.

## 5. RESULTS VERIFICATION AND DISCUSSIONS

Many researchers face difficulty in generating motion files for the movable anchor node for various types of fixed path algorithms. This section demonstrates how the proposed framework solves the problem stated by applying an experimental setup and then verifying the obtained results. With this proposed framework, researchers can make their own scenario files of how the movable anchor node might move based on different static path planning mechanisms. Afterward, the generated scenario file can be used by users in the network simulator (NS-2). In the following, an experimental setup for different static path planning mechanisms (case studies) will be conducted.

### 5.1. SCAN Trajectory Model

The following table shows the experimental setups used for simulating SCAN path planning mechanism.

Table 1. SCAN Experimental Setup

| Input Parameters | Value |
|---|---|
| Number of Horizontal Segments (R) | 10 |
| Trajectory Resolution (m) | 50 |
| Anchor Mobility Speed (m/Sec) | 10 |
| Total Number of Nodes | 10 |

The next figure shows what the proposed framework looked like after 3 minutes of simulation. As shown in Figure 9, the trajectory of the SCAN model has been drawn on the monitoring area as a series of vertical and horizontal segments in black lines, and there are nine sensor nodes as blue (hollow) nodes and the movable anchor shown by an orange (solid) node. There are five nodes: four of them lie at the four corners, and one lies in the middle of the monitoring area. These nodes are called *Base Stations* for negotiation and data dissemination.

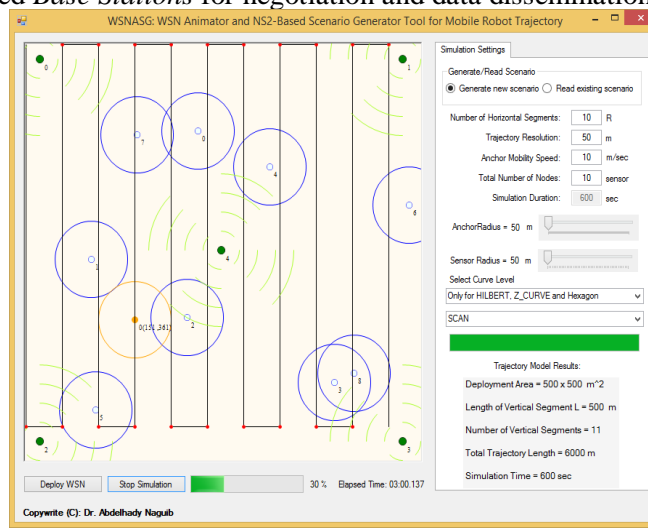

Figure 9. Simulation of SCAN Trajectory Model

The output of this SCAN mechanism is represented in the following table. There are two kinds of generated files: Excel files and text files. Figure 10 represents the chart of the SCAN trajectory model that was sketched using the generated Excel file.

Table 2. Simulation Results of the SCAN Mechanism

| Results | Value |
| --- | --- |
| Area Size (m$^2$) | 500 x 500 |
| Length of Vertical Segment (m) | 500 |
| Number of Vertical Segments | 11 |
| Total Trajectory Length (m) | 6000 |
| Simulation Time (Sec) | 600 |

The chart shown in Figure 10 is considered evidence that the generated SCAN trajectory in Figure 9 is correct. The other text (scenario) file that was made after running the SCAN mechanism is shown in the figure below. This scenario file is used by the NS-2 simulator to simulate how the movable anchor node moves.

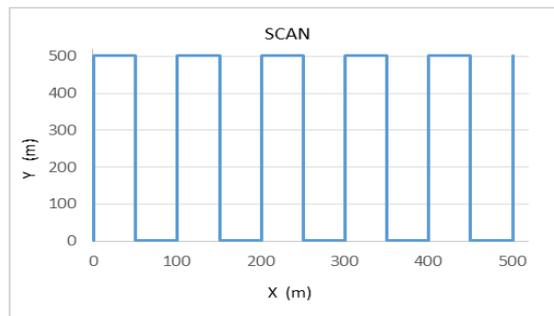

Figure 10. Chart of the SCAN Trajectory

The generated scenario file from the proposed framework in Figure 11 was applied to the NS-2 simulator to ensure that the results of the proposed framework were correct.

Figure 11. SCAN Scenario File

International Journal of Computer Networks & Communications (IJCNC) vol 16, No 6, November 2024

Figure 12 is considered evidence of the correctness of the proposed framework results. It represents a screenshot from NS-2 running the SCAN trajectory. The area is 500 x 500 m$^2$, the length of a vertical segment is 500 m, and there are 11 vertical segments, as shown. These results are compatible with the results gained from the proposed framework in Table 2.

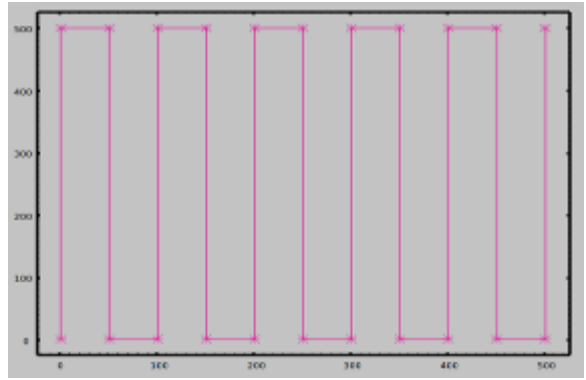

Figure 12. Screenshot of NS-2 Running SCAN Trajectory

## 5.2. HILBERT Trajectory Model

In this experiment, the HILBERT trajectory model has been simulated according to the parameters shown in Table 3.

Table 3. HILBERT Experimental Setup

| Input Parameters | Value |
|---|---|
| Trajectory Resolution (m) | 35 |
| Anchor Mobility Speed (m/Sec) | 10 |
| Total Number of Nodes | 5 |
| Curve Level | 4 |

The shape of the HILBERT path (level 4) shown in Figure 13 is generated after applying the parameters in Table 3. The anchor moves at a speed of 10 m/sec and reaches the middle of the monitoring area after 11 minutes (i.e., 71.0%) of simulation time.

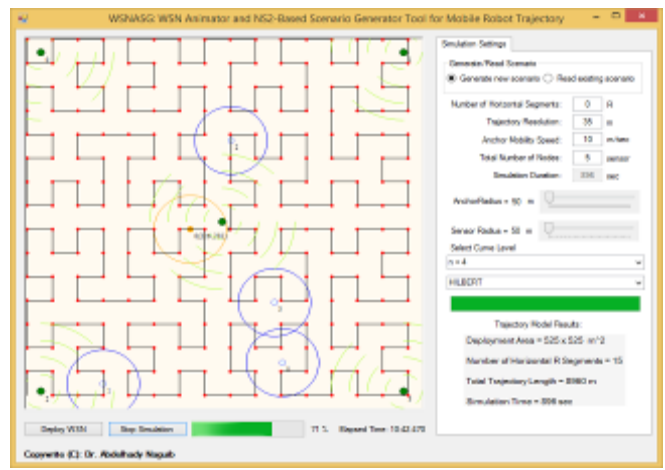

Figure 13. Simulation of the HILBERT Trajectory Model

31

International Journal of Computer Networks & Communications (IJCNC) vol 16, No 6, November 2024

The following table presents the results when simulating the HILBERT mechanism. As shown, the deployment field is 525 x 525 m$^2$, which is approximately equal to the area of the SCAN model, but the total trajectory length of the HILBERT model is 8960 m, which is longer than the path of the SCAN model due to the many turns of the HILBERT model.

Table 4. Simulation Results of the HILBERT Mechanism

| Results | Value |
| --- | --- |
| Area Size (m$^2$) | 525 x 525 |
| Number of Horizontal Segment | 15 |
| Total Trajectory Length (m) | 8960 |
| Simulation Time (Sec) | 896 |

From Figure 14, the generated HILBERT path from the Excel file proves that it is compatible with the trajectory present in Figure 13. The following figure represents the scenario file for the movement of the anchor node, which can be used in the NS-2.

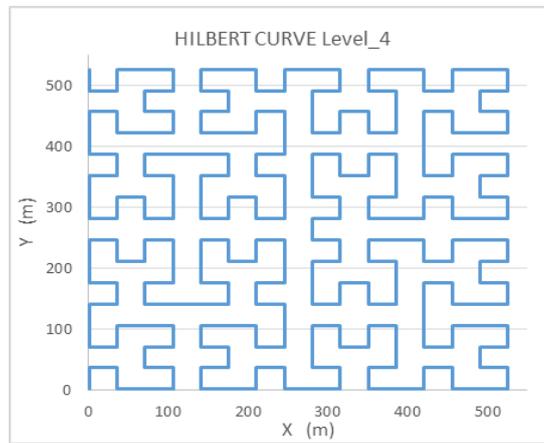

Figure 14. Chart of the HILBERT Trajectory

In order to verify the generated HILBERT scenario file (Figure 15), it has been used by the NS-2, and the resulting trajectory from this simulation is shown in Figure 16. The trajectory represented in Figure 16 is the same as the one in Figure 13.

Figure 15. HILBERT Scenario File

International Journal of Computer Networks & Communications (IJCNC) vol 16, No 6, November 2024

It can be concluded that the output of the proposed framework gives the same output as the NS-2. As shown in Figure 16, the area size is 525 x 525 m² and the number of horizontal segments is 15. Hence, these results are considered evidence for the correctness of the proposed framework results.

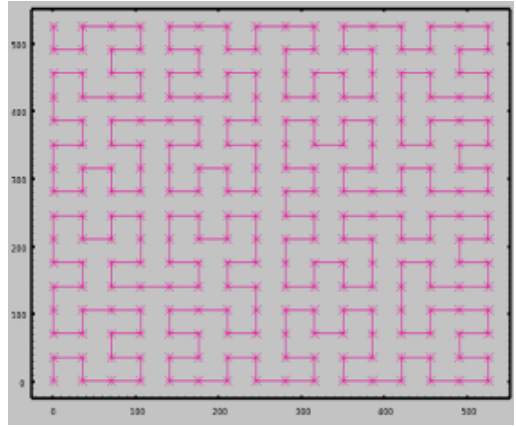

Figure 16. Screenshot of NS-2 Running the HILBERT

## 5.3. SPIRAL Trajectory Model

In this experiment, the SPIRAL trajectory model has been simulated according to the parameters shown in Table 5.

Table 5. SPIRAL Experimental Setup

| Input Parameters | Value |
| --- | --- |
| Number of Horizontal Segments | 10 |
| Trajectory Resolution (m) | 50 |
| Anchor Mobility Speed (m/Sec) | 8 |
| Total Number of Nodes | 6 |

As shown in Figure 17, the movable anchor node reaches coordinate of the point (376, 282) after 2 minutes of simulation at a speed of 8 m/sec.

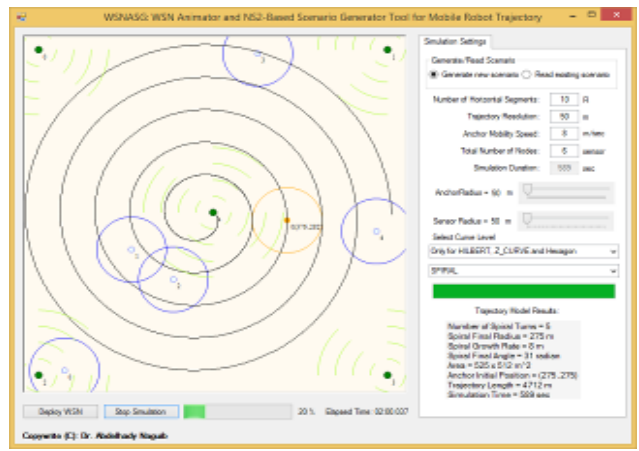

Figure 17. Simulation of the SPIRAL Trajectory Model

33

The results of the SPIRAL trajectory model are shown in Table 6. As shown, the movable anchor node started moving from its initial position in the middle of the monitoring area, and after 589 seconds (the end of the simulation), it traversed a distance of 4712 m. The traversed distance of the movable anchor node for the SPIRAL mechanism is smaller than SCAN and HILBERT, respectively.

Table 6. Simulation Results of the SPIRAL Mechanism

| Results | Value |
| --- | --- |
| Number of Spiral Turns | 5 |
| Spiral Final Radius (m) | 275 |
| Spiral Growth Rate (m) | 8 |
| Spiral Final Angle (rad) | 31 |
| Area Size (m$^2$) | 525 x 512 |
| Anchor Initial Position (x, y) | (275, 275) |
| Total Trajectory Length (m) | 4712 |
| Simulation Time (sec) | 589 |

The chart representing SPIRAL trajectory in Figure 18 shows that the monitoring area size equals 525 x 512 m$^2$, the initial position of the movable anchor is at point (275, 275), and the final radius of SPIRAL trajectory is 275 m. These observations are compatible with the results shown in Table 6.

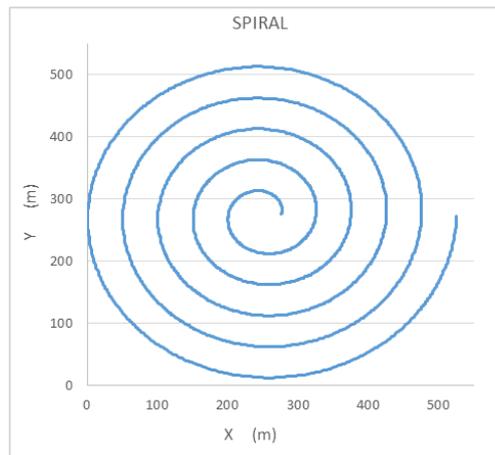

Figure 18. Chart of the SPIRAL Trajectory

The chart of SPIRAL trajectory shown in Figure 18 is the same as the trajectory drawn in the user interface of Figure 17. Note that the origin point in the coordinate system of the proposed framework lies at the upper-left corner of the monitoring area. Hence, the increment in the y-axis goes down differently than in the well-known coordinate system. Therefore, the trajectory in Figure 17 is a reflected image of the trajectory in Figure 18 This notice applies to all trajectories generated from the proposed framework but does not affect the results of the NS-2.

![Figure 19 SPIRAL Scenario File - Notepad screenshot showing ns_ at time commands with node setdest coordinates]

Figure 19. SPIRAL Scenario File

The scenario file for the SPIRALL trajectory shown in Figure 19 contains the coordinates of the movable anchor at certain times. As shown in Figure 19, the difference between two successive points on the SPIRAL trajectory is very small. Thus, the motion of the anchor node gives an exact spiral shape, which is an indication that the proposed framework gives high-accuracy results.

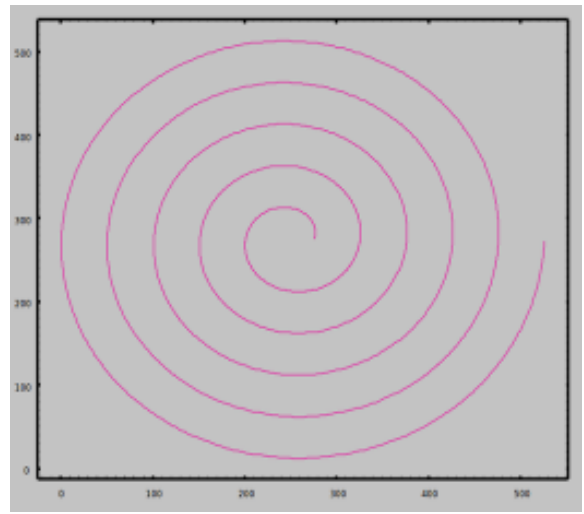

Figure 20. NS-2 Screenshot Running the SPIRAL Trajectory

The SPIRAL shape shown in Figure 20 results from the NS-2 simulator, and as shown, it is compatible with Figures 17 and 18. Hence, it is evidence that the results of the proposed framework have been verified.

## 5.4. Comparing the Efficiency of the Proposed Framework against other Tools

In this section, the novelty of the proposed framework has been clarified. In the following we a brief description of how this framework differs from existing WSN simulation tools. A comparison was made between the proposed framework and related systems [26 – 29]. As shown in Table 7, the proposed framework has a very good design for the user interface but the other tools fall between bad and good. According to robot movement, the proposed framework supports trace files for robot movement, which leads to outstanding performance among other tools. As for the last two metrics: generating scenarios and path models, the proposed framework supports configuration for different scenarios and can validate obtained results by running real time simulations for the movement files of many path planning models that already exist or can be added to the proposed framework.

Table 7. Comparison of the proposed framework against four related systems

| Simulation Tools | User Interface Design | Robot Movement | Generating Scenarios | Path Models |
|---|---|---|---|---|
| Ali, Q. I. [26] | Poor | Good | No support | No support |
| Niewiadomska et al. [27] | Acceptable | Basic | No support | No support |
| Pagano et al. [28] | low | Average | No support | No support |
| Mozumdar et al. [29] | Good | Poor | No support | No support |
| The Proposed Framework | Very Good | Outstanding | Superior | Most models |

As shown in Table 8, a quantitative evaluation is conducted based on eight different features this is to prove that the proposed model outperforms its other related tools. As shown, the proposed framework is an extendable system that supports adding new models to its platform as needed. The proposed framework supports interaction design which improves usability makes it easier and more intuitive to use and enhances user satisfaction. Flexibility was taken into account when designing the framework which means the system is able to adapt to changing requirements and technologies. The operability feature is supported by the proposed framework, meaning that the users can interact easily with the framework GUI and it has a good operability works. The proposed framework supports validity, ensuring that the framework meets users' expectations. Scalability feature is considered one of the most important characteristics of the framework which eliminates the burden of limiting future operations. Framework reliability is an essential feature of system quality ensuring it is free from failure. Finally, the simplicity of the proposed framework can significantly impact efficiency and performance making it easier to maintain, debug, and enhance the system over time. As shown in Table 8, the proposed framework outperforms related tools in all necessary features for any reliable system.

Table 8. Evaluation of the proposed framework based on eight features

| Feature | Ali, Q. I. [26] | Niewiadomska et al. [27] | Pagano et al. [28] | Mozumdar et al. [29] | The Proposed Framework |
|---|---|---|---|---|---|
| Extendibility | ✗ | ✗ | ✓ | ✓ | ✓ |
| Interaction | ✓ | ✓ | ✗ | ✗ | ✓ |
| Flexibility | ✓ | ✗ | ✓ | ✗ | ✓ |
| Operability | ✗ | ✗ | ✗ | ✗ | ✓ |
| Validity | ✓ | ✗ | ✓ | ✓ | ✓ |
| Scalability | ✗ | ✓ | ✗ | ✓ | ✓ |
| Reliability | ✓ | ✓ | ✓ | ✓ | ✓ |
| Simplicity | ✓ | ✗ | ✗ | ✗ | ✓ |

## 6. CONCLUSIONS AND FUTURE WORK

One difficulty that users face when performing simulations using NS-2 is how to create anchor motion files based on different path models. Therefore, this article proposes an outstanding framework for simulating anchor paths in wireless sensor networks and helping researchers not only generate different motion files but also modify them to simulate other static path algorithms. In this article, an introduction is made to the latest and most important static path planning mechanisms in WSNs that have been implemented in the proposed framework. There are several stages during the construction and implementation of the proposed framework, starting with analysis, design, experimental setup, and finally, verification. Broad simulations using the NS-2 have been done to verify the results, and the validation process proved the validity and accuracy of the results of the proposed framework. The proposed framework is an open source system that is available for download from the Code Project website. The proposed framework was compared with existing tools and the results proved the validity of the design and its superiority to them.

According to the future work, this work is currently underway to update the proposed framework to generate motion files recording the anchor's trajectory after avoiding obstacles of a variable size. These obstacles can be changed depending on the facilities given to the user. Also, the proposed framework is a prelude to future work, which is the development and implementation of a WSN localization simulator.

## 7. CONFLICTS OF INTEREST

The authors declare no conflict of interest.

**AUTHOR**

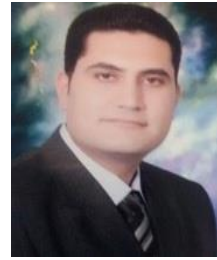

**Abdelhady Naguib** received his M.S. and Ph.D. in Systems and Computers Engineering from Al-Azhar University in 2008 and 2013 respectively. He is currently an associate professor at, Faculty of Engineering at Al-Azhar University, Cairo, Egypt. In addition, he is working as assistant professor at department of Computer Science, Jouf University, Kingdom of Saudi Arabia.

He has a research and teaching experience for many years. He is also acting as a reviewer of well-reputed journals like Springer Nature and Inderscience Online. His research area is in the field of Mobile Communications more specific in the area of mobile ad hoc and wireless sensor networks. He has contributed to this area by many papers and one book, "Extending NS-2 for Simulating a Localization Algorithm," 1st. ed., LAP LAMBERT Academic Publishing, April 2001, ISBN-13 978-3-330-07674-7. Dr. Abdelhady Naguib is a member at Egyptian Syndicate for Engineers and has a membership of Cisco academy (as instructor).